# CCD photometry of the first observed superoutburst of KP Cassiopeiae in 2008 October

David Boyd, Pierre dePonthiere, Jerry Foote, Mack Julian, Taichi Kato, Robert Koff, Tom Krajci, Gary Poyner, Jeremy Shears, Bart Staels

## Abstract

We report CCD photometry and analysis of the first observed superoutburst of the SU UMa-type dwarf nova KP Cassiopeiae during 2008 October. We observed a distinct shortening of the superhump period at superhump cycle 15. Before that point $P_{sh}$ was 0.08556(3) d and afterwards it evolved from 0.08517(2) d to 0.08544(3) d with a rate of period change $dP_{sh}/dt = 3.2(2) * 10^{-5}$. We measured the likely orbital period as 0.0814(4) d placing KP Cas just below the period gap. The superhump period excess ε is 0.048(5) and, empirically, the mass ratio q is 0.20(2). The superoutburst lasted between 8 and 12 days, peaked close to magnitude 13 with an amplitude above quiescence of 5 magnitudes, and faded for 4 days at a rate of 0.14 mag/d. Close monitoring following the end of the superoutburst detected a single normal outburst 60 days later which reached magnitude 14.7 and lasted less than 3 days.

## What did we know about KP Cassiopeiae?

According to IBVS 4896 [1], the variable we now refer to as KP Cas was first described by Hoffmeister in 1949 [2] and appeared on MVS chart N291 published by Hoffmeister in 1957 [3] with the name Sonneberg variable S 3865. This chart is available online [4]. Kinnunen and Skiff [1] identified the variable with an 18th magnitude very blue star in the USNO-A2.0 catalogue at position 0h 38m 54s.70 +61° 12' 59".9 (+/-0".5) (J2000) with colour index (b-r) = -0.2. They comment that identifications prior to this had been incorrect. They also report that the variable was recorded in outburst at magnitude 15.5-16 on POSS-II J plate (1989 September 1). The only previous outburst detected visually was by Kinnunen on 1997 September 26 at magnitude 15.6. A search of the American Association of Variable Star Observers (AAVSO) variable star database [5] found about 30 negative observations with a fainter-than magnitude of typically 15.0 during the 6 years prior to the current outburst. With such low coverage, including long periods with no reported observations, it is likely that previous outbursts have been missed.

KP Cas is catalogued in the General Catalogue of Variable Stars [6] as a UGSS-type dwarf nova with a range 15.2p to 20p, while Simbad [7] lists the range as 15.2p to >17.5p. A search on VizieR [8] found entries for KP Cas in the USNO-A2.0, USNO-B1.0 and GSC 2.3 catalogues with B, V, and R magnitudes in the range 17.6 to 18.8. We will adopt 18th magnitude as the quiescent level of KP Cas for the purpose of this report.

## This outburst

The outburst described here was first detected by Yasuo Sano (Hokkaido, Japan) on 2008 October 25.534 (JD 2454765.034) and reported on a VSNET email list [9]. He recorded KP Cas at magnitude 13.1C using a 0.3-m f/4 reflector and BJ42L CCD camera while four days previously he had recorded nothing brighter than 15.5C at that location. The outburst was confirmed visually by Eddy Muyllaert (Oostende, Belgium) on October 27.760 at magnitude 14.0 and by Poyner on October 27.771 at magnitude 13.9.

We began time series CCD photometry on October 27 (JD 2454767) and immediately recorded superhumps confirming that KP Cas was a SU UMa-type dwarf nova and this was the first observed superoutburst. We recorded a total of 144 hours of data comprising 8412 observations on 8 of the next 11 days. Observations were generally made with either a V filter or unfiltered (C). The log of time series observations and the equipment used are listed in Tables 1 and 2 respectively. Figure 1 shows an image of the field of KP Cas taken on October 27.

**Photometry**

All images were dark-subtracted and flat-fielded and instrumental magnitudes obtained using aperture photometry. Comparison stars were selected from the KP Cas field calibration data provided by Arne Henden from observations made at the Sonoita Research Observatory [10]. Selection criteria for comparison stars included as similar as possible magnitude and colour to KP Cas, proximity to KP Cas for field of view considerations, and separation from adjacent stars which might contaminate photometry. All selected stars were checked for constancy using photometry taken over several nights and one nearby bright star originally included as a comparison was eliminated because of low amplitude variability. This new variable is marked in Figure 1 and will be reported on in a separate paper. The comparison stars selected are listed in Table 3.

A light curve of the superoutburst is shown in Figure 2 including all our time series observations and the 3 early observations noted above. The initial observation by Sano appears slightly bright compared to the later trajectory of the light curve but this may indicate that the first stage of the decline was more rapid. The beginning of the outburst was not observed but from Sano's negative observation 4 days previously we know it occurred after JD 2454761.0. Assuming the quiescent level of KP Cas is ~18$^{th}$ magnitude, this superoutburst had an amplitude close to 5 magnitudes and lasted between 8 and 12 days, normal for an SU UMa-type dwarf nova. From the start of our time series observations, the outburst declined at 0.14 mag/d for 4 days before fading more rapidly towards quiescence.

V and R-band photometry obtained on October 27 near the peak of the outburst gave colour index (V-R) = 0.13(2) and on November 6 with Faulkes Telescope North (FTN) when KP Cas was at magnitude 17.85V gave (V-R) = 0.45(19). The large error on the FTN measurement is due to the low S/N ratio of the KP Cas image on the FTN 60 sec V and R exposures.

**Astrometry**

Astrometry of KP Cas on 10 images obtained under good conditions on JD 2454767 using Astrometrica [11] and the USNO-B1.0 catalogue gave a mean position for KP Cas of 0h 38m 54s.82 +61° 13' 0".5 +/- 0".2 (J2000).

**Superhump timing analysis**

Superhump evolution during the outburst is illustrated in Figure 3 which shows a selection of light curve segments, all at the same scale. These segments are aligned in superhump phase and have the same vertical magnitude scale. The vertical scale mark represents 0.1 magnitude. There is a slow decrease in amplitude from ~0.25 magnitude on JD 2454767 to ~0.15 magnitude on JD2454771 but the final time series observation on JD 2454776 shows a large increase in amplitude and also the development of a secondary hump between the primary superhumps, possibly indicating the presence of the first harmonic of the superhump signal.

During the outburst 41 superhumps were sufficiently well observed by European and American observers that a quadratic fit to the light curve of each superhump provided a reliable measurement of its time of maximum and associated error. Timings for 39 superhumps obtained by Japanese observers were provided by Kato. These timings were obtained by fitting a standard superhump template to each superhump. Comparing timings obtained for the same superhumps using both methods revealed a small systematic difference which was compensated for by adjusting the timings from Kato by –0.0017 d. Some superhumps were observed by more than one observer. Heliocentric corrections were applied to all times of maximum.

A preliminary linear fit to all times of maximum enabled unambiguous superhump cycle numbers to be assigned to each superhump. Using weights equal to the inverse square of the timing errors, a weighted linear fit to these timings and cycle numbers gives the superhump maximum ephemeris

$$HJD(max) = 2454767.0253(1) + 0.085292(3) * E \quad (1)$$

The O-C (observed minus calculated) times relative to this ephemeris are plotted in Figure 4. This shows a distinct shortening in the superhump period $P_{sh}$ around superhump cycle 15 (JD 2454768.3). Before cycle 15, $P_{sh}$ appears to be constant at 0.08556(3) d with linear ephemeris

$$HJD(max) = 2454767.0231(2) + 0.08556(3) * E \quad (2)$$

Table 4 gives superhump timings and O-C values relative to eqn (2). At about cycle 15, $P_{sh}$ suddenly shortens and then over the cycle range 16-115 progressively lengthens. We assume the superhumps at cycles 113-115 (JD 2454776) are part of this progression. The superhump timings in this region follow the quadratic ephemeris

$$HJD(max) = 2454767.0293(3) + 0.08513(1) * E + 0.00000137(9) * E^2 \quad (3)$$

Table 5 gives superhump timings and O-C values relative to eqn (3). The mean value of $P_{sh}$ over this region is 0.085274(3) d while at the beginning it is 0.08517(2) d and at the end 0.08544(3) d. The mean rate of change of $P_{sh}$ over this interval is $dP_{sh}/dt = 3.2(2) * 10^{-5}$. Positive superhump period derivatives are unusual for systems with such a large value of $P_{sh}$ [12, 13] but this may be a consequence of analysing different stages of a superoutburst. To illustrate this point, analysing the data for cycles 0-65 gives a negative rate of change of $P_{sh}$ with $dP_{sh}/dt = -7.1(5) * 10^{-5}$ which is consistent with values reported in [12] and [13].

**Period analysis**

After subtracting linear trends from each dataset we performed a Lomb-Scargle [14,15] period analysis on all our time series data using Peranso [16]. The resulting power spectrum is shown in Figure 5. The strongest signal is at 0.0853(2) d (11.72(2) c/d) and there are low amplitude 1 and 2 c/d alias signals around this as expected from a spectral window analysis of the data. This signal is consistent with the mean value of $P_{sh}$ derived from the superhump timing analysis. There is also a low power first harmonic of the superhump signal at 0.04266(4) d (23.44(2) c/d) as expected from the appearance of secondary humps in the superhump light curve towards the end of the outburst. Prewhitening to remove the fundamental superhump signal gives the power spectrum in Figure 6. The residual signals 0.1 c/d above and below the fundamental superhump frequency are probably due to incomplete removal of the superhump signal because of its change in frequency during the outburst. Just above the superhump frequency is a signal at 0.0814(4) d (12.28(6) c/d) with no matching alias signal below the superhump frequency which would be expected if it was an alias of the superhump signal. We suggest this signal is the orbital period of KP Cas which, at 1.95 hours, puts KP Cas just below the dwarf nova period gap.

We also performed separate Lomb-Scargle analyses on the time series data before and after cycle 15, the point at which $P_{sh}$ changed. The superhump periods from these analyses are given in Table 6 along with the results of the superhump timing analysis. Both methods gave consistent results. Phase diagrams of the superhump signal folded on the mean values of $P_{sh}$ in Table 6 are shown in Figure 7 for the cycles ranges 0-15, 16-65 and 113-115. The mean superhump amplitudes in these cycle ranges are 0.23, 0.17 and 0.24 magnitudes .

**Discussion**

The apparent regrowth in superhump amplitude which we see in the data during JD 2454776 has been seen in other SU UMa-type dwarf novae [17] but has tended to occur in shorter period systems and during the superoutburst rather than after the end of the outburst as seen here. However, given our lack of observations in the preceding 3 days, we cannot say when this regrowth started.

The phase diagram for cycles 113-115 shows evidence of the first harmonic signal and this is confirmed by the presence of a strong signal at this frequency in a period analysis of this dataset alone. In V1316 Cyg [18], the transition from common to late superhumps was associated with the appearance of the first harmonic of the superhump signal so it is possible that around cycle 115 there was a transition to late superhumps which would have had a shorter period. However, as these were not observed, this can only be speculation.

From the mean superhump period of 0.085292(3) d and the orbital period of 0.0814(4)d, we find a superhump period excess $\varepsilon = 0.048(5)$, in line with results for other SU UMa-type dwarf novae with similar orbital periods [19]. If we assume KP Cas is a normal SU UMa-type dwarf nova with a white dwarf mass of ~0.75 solar masses, then using the empirical relationship $\varepsilon = 0.18*q + 0.29*q^2$ from [20] gives the secondary to primary mass ratio q = 0.20(2).

**Subsequent normal outburst**

Following the superoutburst, Shears monitored KP Cas closely. Surprisingly, given the long interval since the last recorded outburst of KP Cas, he detected another outburst on 2009 January 4 only sixty days after the end of the superoutburst. From fainter than magnitude 17.3C on January 3.806, it reached 15.5C on January 4.778, 14.6C on January 5.758 and by January 9.832 had returned to 17.7C. Time series photometry over 5 hours by Roger Pickard, Shears and Boyd on January 5 found it fading from magnitude 14.6C at a rate of 1 mag/d with no significant modulation. With this rate of decline near the peak of the outburst, we estimate that the duration of the outburst was probably less than 3 days. Its shorter duration, lower amplitude and lack of modulation indicate this was a normal outburst.

## Future monitoring

Previous observations of KP Cas have been too infrequent to provide any useful information about its outburst frequency. Future monitoring of this variable, which has relatively bright outbursts, is required if we are to learn more about its outburst behaviour.

## Conclusion

We report CCD photometry and analysis of the first observed superoutburst of the SU UMa-type dwarf nova KP Cas during 2008 October. Measurements of times of superhump maximum revealed a distinct shortening of the superhump period $P_{sh}$ at cycle 15. Before that point $P_{sh}$ was 0.08556(3) d and afterwards it evolved from 0.08517(2) d to 0.08544(3) d with a rate of period change $dP_{sh}/dt = 3.2(2) * 10^{-5}$. Frequency analysis of our time series data revealed the likely orbital period of KP Cas to be 0.0814(4) d (1.95 h). This places it just below the dwarf nova period gap. The superhump period excess $\varepsilon$ is 0.048(5) and, empirically, we deduce the mass ratio q to be 0.20(2). The outburst lasted between 8 and 12 days, peaked close to magnitude 13, had an amplitude above quiescence of 5 magnitudes, and declined for 4 days at a rate of 0.14 mag/d before fading more rapidly back to quiescence. This is normal superoutburst behaviour for a SU UMa-type dwarf nova. 60 days after the end of the superoutburst we detected a single normal outburst which reached magnitude 14.7 and lasted less than 3 days.

## Acknowledgements

We acknowledge with thanks VSNET for its announcement of the discovery, the AAVSO for variable star observations from its International Database contributed by observers worldwide and used in this research, the NASA Astrophysics Data System and the Simbad and VizieR services operated by CDS Strasbourg. We are grateful to Wolfgang Renz for his help in researching historical information on KP Cas and to Professor Joe Patterson for agreeing to the use of data submitted to the Centre for Backyard Astrophysics. We thank Roger Pickard for his data on the subsequent normal outburst. We also thank the referees for their helpful comments.

## Addresses

DB: 5 Silver Lane, West Challow, Wantage, Oxon, OX12 9TX, UK [drsboyd@dsl.pipex.com]

PdeP: 15 rue Pre Mathy, 5170 Lesve-Profondeville, Belgium [pierredeponthiere@gmail.com]

JF: CBA Utah, 4175 East Red Cliffs Drive, Kanah, UT 84741, USA [jfoote@scopecraft.com]

MJ: 4587 Rockaway Loop, Rio Rancho, NM 871224, USA [mack-julian@cableone.net]

TKa: Department of Astronomy, Kyoto University, Kyoto 606-8502, Japan [tkato@kusastro.kyoto-u.ac.jp]

RK: CBA Colorado, 980 Antelope Drive West, Bennett, CO 80102, USA [bob@antelopehillsobservatory.org]

TKr: CBA New Mexico, PO Box 1351 Cloudcroft, New Mexico 88317, USA [tom_krajci@tularosa.net]
GP: 67 Ellerton Road, Kingstanding, Birmingham B44 0QE, UK [garypoyner@blueyonder.co.uk]
JS: "Pemberton", School Lane, Bunbury, Tarporley, Cheshire, CW6 9NR, UK [bunburyobservatory@hotmail.com]
BS: CBA Flanders (Alan Guth Observatory), Koningshofbaan 51, B-9308 Hofstade, Belgium [staels.bart.bvba@pandora.be]

| Start time (JD) | Duration (hrs) | Filter | Observer |
|---|---|---|---|
| 2454767.26039 | 7.70 | V+R | Boyd |
| 2454767.36072 | 4.10 | C | Shears |
| 2454767.61938 | 8.13 | V | Julian |
| 2454768.24024 | 7.14 | C | Shears |
| 2454768.26340 | 6.40 | C | Staels |
| 2454768.57255 | 10.48 | V | Koff |
| 2454769.22919 | 4.17 | C | dePonthiere |
| 2454769.23649 | 3.66 | C | Staels |
| 2454769.54403 | 10.60 | V | Koff |
| 2454769.60627 | 8.44 | V | Julian |
| 2454770.27623 | 8.24 | C | Boyd |
| 2454770.28947 | 5.77 | C | Shears |
| 2454770.53278 | 11.01 | V | Koff |
| 2454770.60228 | 8.54 | V | Julian |
| 2454771.23582 | 7.97 | C | Boyd |
| 2454771.26962 | 7.00 | C | Shears |
| 2454771.65566 | 7.56 | C | Krajci |
| 2454772.23536 | 1.56 | C | Shears |
| 2454772.42193 | 7.35 | C | dePonthiere |
| 2454776.58473 | 1.27 | C | Krajci |
| 2454776.63559 | 6.74 | C | Foote |
| 2454777.00235 | 0.04 | V+R | Boyd |

Table 1. Log of time series observations.

| Observer | Equipment used |
|---|---|
| Boyd | 0.35-m f/5.3 SCT + SXV-H9 CCD<br>2.0-m f/10 Ritchey-Chretien + E2V-4240 CCD (FTN) |
| dePonthiere | 0.2-m f/6.3 SCT + ST-7XMEI CCD |
| Foote | 0.60-m f/3.4 reflector + ST-8e CCD |
| Julian | 0.30-m f/10 SCT + SBIG ST10XME CCD |
| Koff | 0.25-m f/10 SCT + Apogee AP-47 CCD |
| Krajci | 0.28-m f/10 SCT + ST-7E CCD |
| Shears | 0.1-m fluorite refractor + SXV-M7 CCD<br>0.28-m f/6.3 SCT + SXVF-H9 CCD |
| Staels | 0.28-m f/6.3 SCT + MX-716 CCD |

Table 2. Equipment used.

| RA (J2000) | Dec (J2000) | V | dV | (B-V) | (V-R) |
|---|---|---|---|---|---|
| 0 39 08.88 | +61 12 03.3 | 12.841 | 0.007 | 0.329 | 0.209 |
| 0 39 16.02 | +61 13 50.1 | 13.955 | 0.009 | 0.435 | 0.255 |
| 0 38 59.76 | +61 13 52.7 | 14.487 | 0.008 | 0.468 | 0.292 |

Table 3. Comparison stars used.

| Superhump cycle no | Observed time of maximum (HJD) | Uncertainty (day) | O-C (day) |
|---|---|---|---|
| 0 | 2454767.0239 | 0.0003 | 0.0008 |
| 1 | 2454767.1083 | 0.0004 | -0.0004 |
| 3 | 2454767.2798 | 0.0005 | 0.0000 |
| 3 | 2454767.2810 | 0.0005 | 0.0012 |
| 4 | 2454767.3643 | 0.0005 | -0.0011 |
| 4 | 2454767.3583 | 0.0014 | -0.0071 |
| 5 | 2454767.4507 | 0.0004 | -0.0002 |
| 5 | 2454767.4507 | 0.0008 | -0.0002 |
| 5 | 2454767.4534 | 0.0014 | 0.0025 |
| 6 | 2454767.5354 | 0.0004 | -0.0010 |
| 6 | 2454767.5423 | 0.0012 | 0.0058 |
| 8 | 2454767.7071 | 0.0004 | -0.0005 |
| 8 | 2454767.7079 | 0.0005 | 0.0003 |
| 9 | 2454767.7922 | 0.0006 | -0.0009 |
| 9 | 2454767.7928 | 0.0005 | -0.0004 |
| 10 | 2454767.8785 | 0.0005 | -0.0003 |
| 10 | 2454767.8781 | 0.0005 | -0.0007 |
| 11 | 2454767.9651 | 0.0006 | 0.0008 |
| 12 | 2454768.0495 | 0.0006 | -0.0004 |
| 14 | 2454768.2211 | 0.0011 | 0.0001 |
| 15 | 2454768.3068 | 0.0006 | 0.0002 |
| 15 | 2454768.3077 | 0.0004 | 0.0011 |

Table 4. Times of superhump maximum and O-C values relative to the linear ephemeris in eqn (2).

| Superhump cycle no | Observed time of maximum (HJD) | Uncertainty (day) | O-C (day) |
|---|---|---|---|
| 16 | 2454768.3916 | 0.0007 | -0.0001 |
| 16 | 2454768.3936 | 0.0005 | 0.0019 |
| 16 | 2454768.3913 | 0.0005 | -0.0004 |
| 17 | 2454768.4766 | 0.0007 | -0.0002 |
| 17 | 2454768.4779 | 0.0005 | 0.0011 |
| 17 | 2454768.4768 | 0.0002 | -0.0001 |
| 18 | 2454768.5606 | 0.0003 | -0.0014 |
| 19 | 2454768.6458 | 0.0011 | -0.0015 |
| 19 | 2454768.6464 | 0.0025 | -0.0008 |
| 20 | 2454768.7327 | 0.0011 | 0.0003 |
| 21 | 2454768.8167 | 0.0011 | -0.0009 |
| 22 | 2454768.9034 | 0.0007 | 0.0006 |
| 23 | 2454768.9870 | 0.0012 | -0.0010 |
| 23 | 2454768.9886 | 0.0007 | 0.0006 |
| 24 | 2454769.0757 | 0.0014 | 0.0026 |
| 27 | 2454769.3288 | 0.0006 | 0.0001 |
| 27 | 2454769.3287 | 0.0028 | -0.0001 |
| 27 | 2454769.3287 | 0.0003 | 0.0000 |
| 30 | 2454769.5840 | 0.0011 | -0.0004 |
| 31 | 2454769.6698 | 0.0009 | 0.0002 |
| 31 | 2454769.6695 | 0.0005 | -0.0001 |
| 32 | 2454769.7555 | 0.0013 | 0.0007 |
| 32 | 2454769.7558 | 0.0009 | 0.0010 |
| 32 | 2454769.7564 | 0.0005 | 0.0016 |
| 33 | 2454769.8409 | 0.0018 | 0.0009 |
| 33 | 2454769.8412 | 0.0009 | 0.0012 |
| 33 | 2454769.8418 | 0.0005 | 0.0018 |
| 34 | 2454769.9251 | 0.0012 | -0.0001 |
| 34 | 2454769.9256 | 0.0010 | 0.0004 |
| 34 | 2454769.9276 | 0.0008 | 0.0024 |
| 35 | 2454770.0090 | 0.0008 | -0.0014 |
| 36 | 2454770.0950 | 0.0006 | -0.0007 |
| 37 | 2454770.1832 | 0.0013 | 0.0023 |
| 39 | 2454770.3511 | 0.0006 | -0.0002 |
| 39 | 2454770.3512 | 0.0004 | -0.0001 |
| 39 | 2454770.3510 | 0.0004 | -0.0003 |
| 40 | 2454770.4373 | 0.0006 | 0.0007 |
| 40 | 2454770.4375 | 0.0004 | 0.0009 |
| 40 | 2454770.4372 | 0.0004 | 0.0006 |
| 41 | 2454770.5218 | 0.0004 | 0.0000 |
| 41 | 2454770.5215 | 0.0004 | -0.0003 |
| 41 | 2454770.5216 | 0.0002 | -0.0002 |
| 42 | 2454770.6068 | 0.0008 | -0.0002 |
| 42 | 2454770.6083 | 0.0014 | 0.0012 |
| 42 | 2454770.6056 | 0.0004 | -0.0015 |
| 43 | 2454770.6937 | 0.0013 | 0.0014 |
| 43 | 2454770.6933 | 0.0010 | 0.0010 |
| 43 | 2454770.6921 | 0.0004 | -0.0002 |
| 44 | 2454770.7782 | 0.0014 | 0.0006 |
| 44 | 2454770.7774 | 0.0009 | -0.0002 |
| 44 | 2454770.7767 | 0.0005 | -0.0009 |
| 45 | 2454770.8637 | 0.0012 | 0.0009 |
| 45 | 2454770.8638 | 0.0006 | 0.0010 |
| 45 | 2454770.8621 | 0.0005 | -0.0007 |
| 46 | 2454770.9486 | 0.0012 | 0.0005 |
| 46 | 2454770.9483 | 0.0010 | 0.0002 |
| 46 | 2454770.9471 | 0.0009 | -0.0010 |
| 50 | 2454771.2888 | 0.0006 | -0.0003 |
| 50 | 2454771.2897 | 0.0003 | 0.0006 |
| 50 | 2454771.2890 | 0.0004 | -0.0001 |
| 51 | 2454771.3746 | 0.0005 | 0.0003 |
| 51 | 2454771.3745 | 0.0004 | 0.0001 |
| 51 | 2454771.3738 | 0.0005 | -0.0006 |
| 52 | 2454771.4590 | 0.0006 | -0.0006 |
| 52 | 2454771.4596 | 0.0003 | 0.0000 |
| 52 | 2454771.4577 | 0.0004 | -0.0019 |
| 53 | 2454771.5456 | 0.0006 | 0.0007 |
| 53 | 2454771.5445 | 0.0007 | -0.0004 |
| 53 | 2454771.5467 | 0.0038 | 0.0018 |
| 55 | 2454771.7155 | 0.0004 | 0.0001 |
| 56 | 2454771.8016 | 0.0005 | 0.0008 |
| 57 | 2454771.8868 | 0.0005 | 0.0008 |
| 64 | 2454772.4857 | 0.0024 | 0.0026 |
| 65 | 2454772.5669 | 0.0028 | -0.0015 |
| 113 | 2454776.6661 | 0.0007 | -0.0001 |
| 114 | 2454776.7521 | 0.0005 | 0.0005 |
| 115 | 2454776.8362 | 0.0008 | -0.0008 |

Table 5. Times of superhump maximum and O-C values relative to the quadratic ephemeris in eqn (3).

| Superhump cycle range | HJD range | $P_{sh}$ from superhump timing analysis (d) | $P_{sh}$ from Lomb-Scargle analysis (d) |
|---|---|---|---|
| 0 - 15 | 2454767.0 - 2454768.35 | 0.08556(3) | 0.0856(6) |
| 16 - 115 | 2454768.35 – 2454777.0 | 0.085274(3) | 0.08527(7) |

Table 6. Mean values of $P_{sh}$ before and after superhump cycle 15.

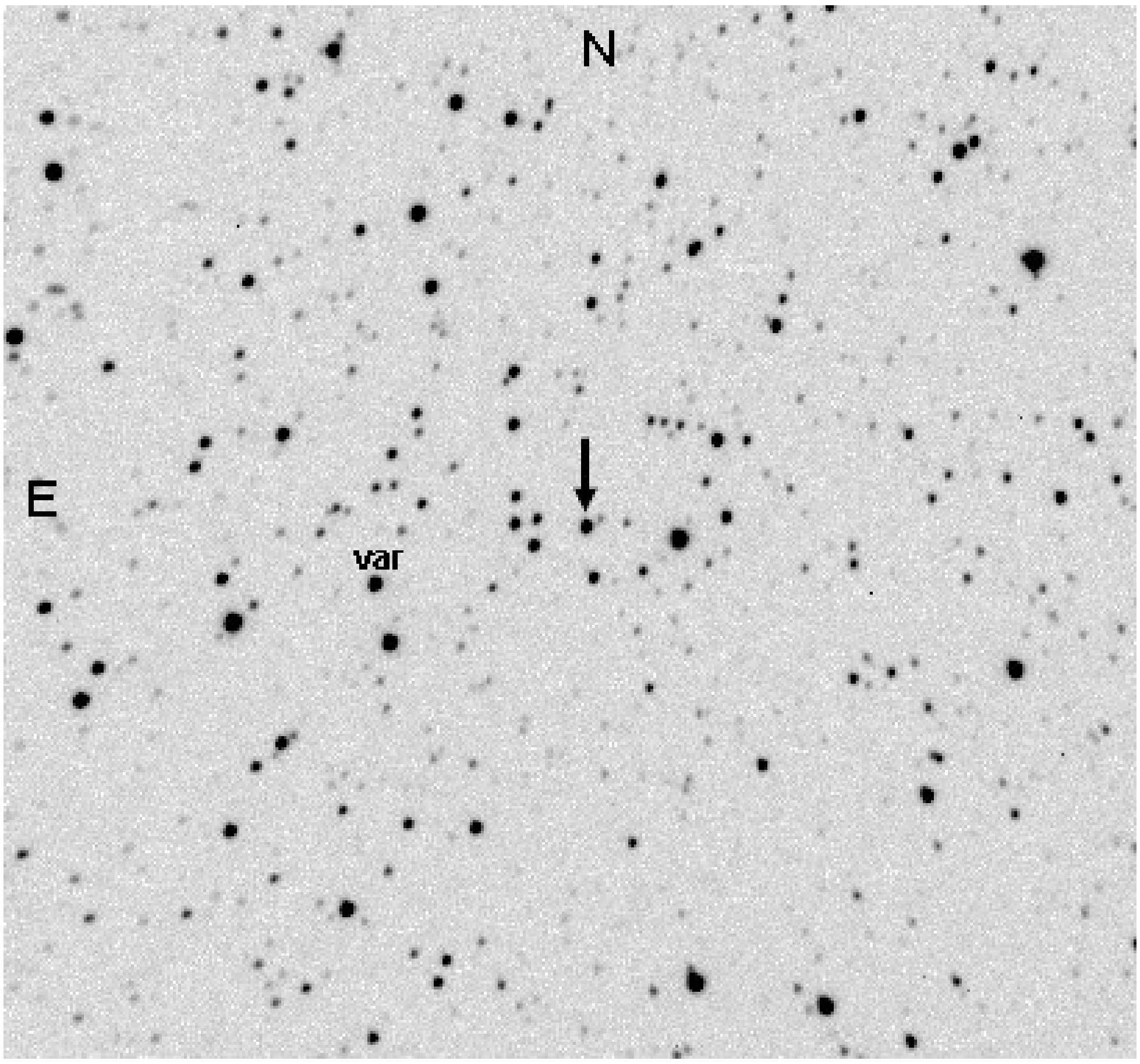

Figure 1. Image of KP Cas taken on October 27, field ~10x10 arcmin. (Boyd)

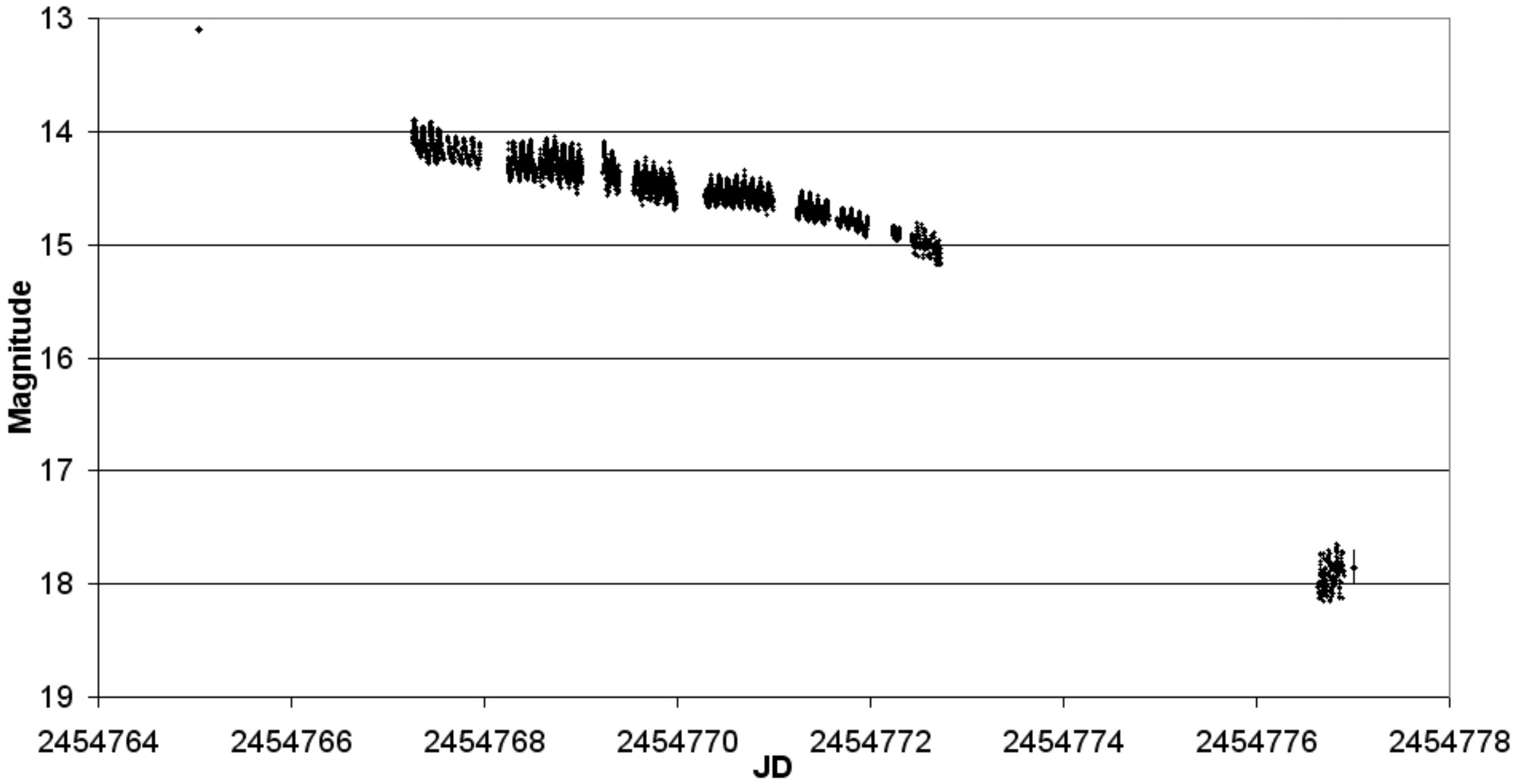

Figure 2. Light curve of the KP Cas superoutburst.

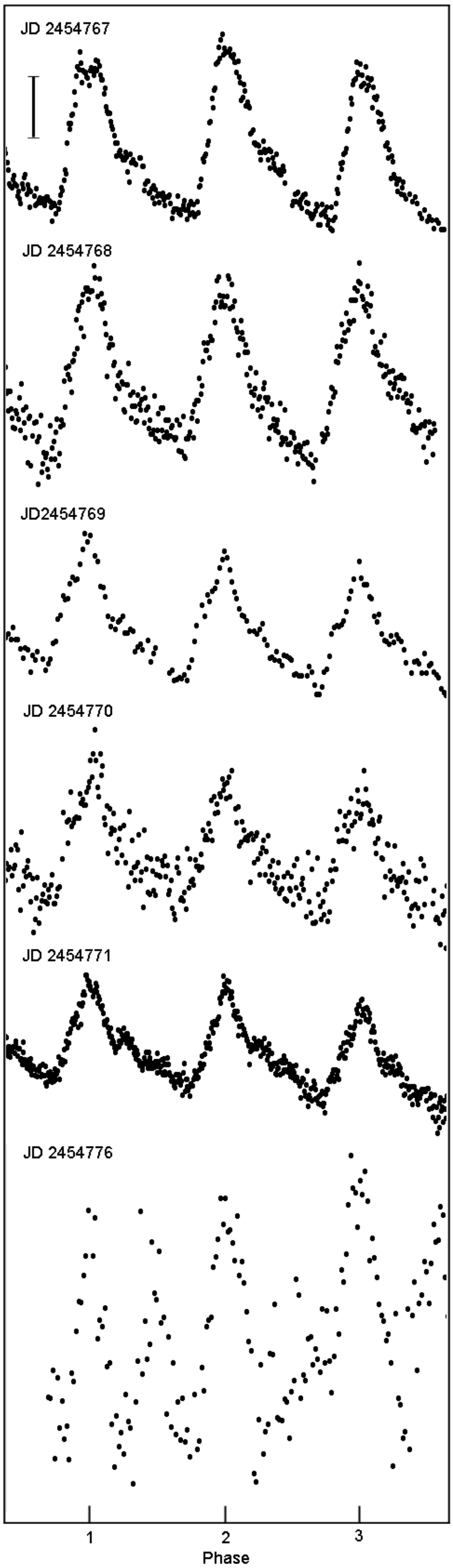

Figure 3. Superhump evolution during the outburst with time shown horizontally and magnitude vertically. All light curves are shown at the same scale in time and magnitude and are aligned in superhump phase. The vertical scale mark is 0.1 magnitude.

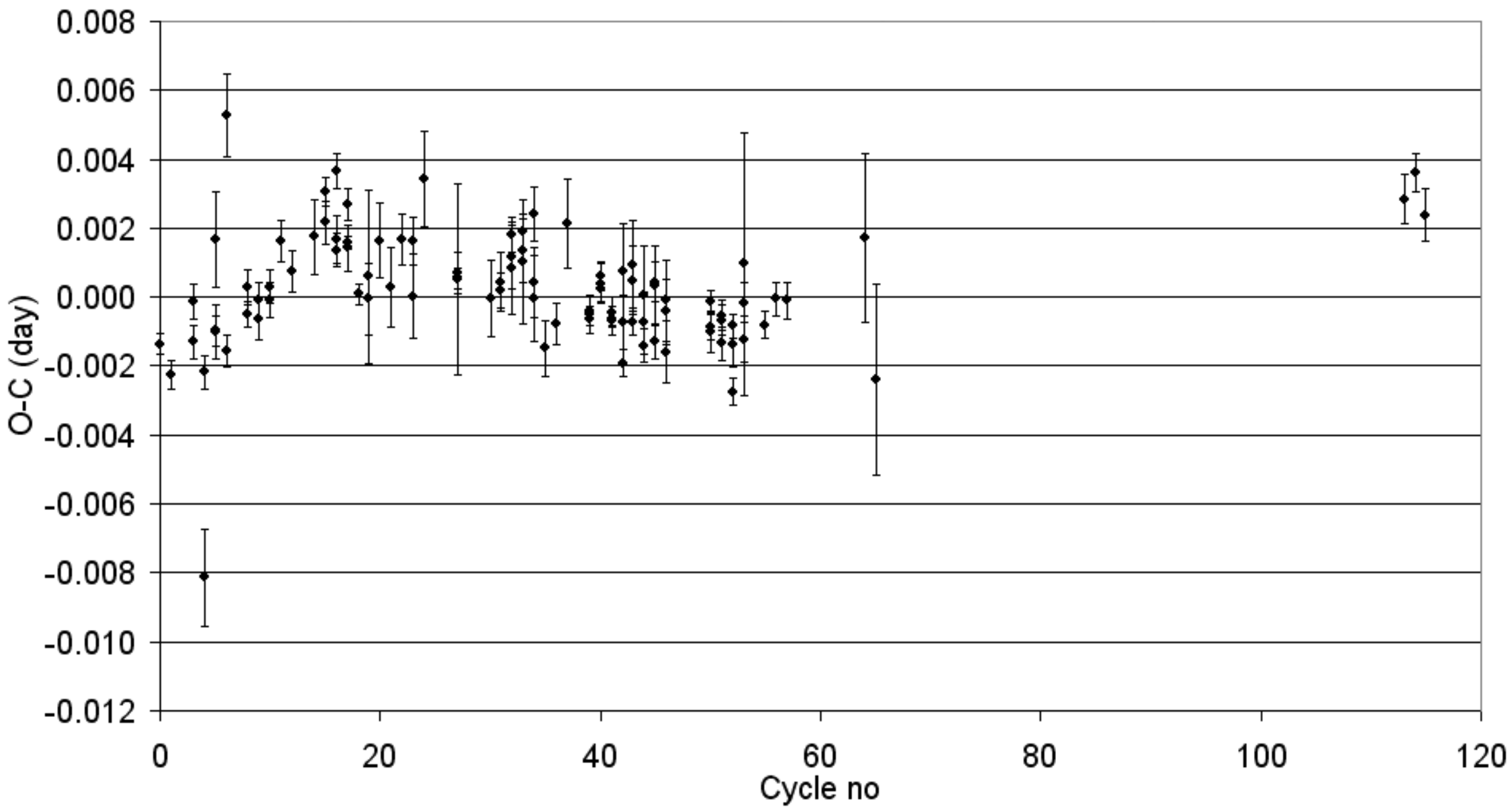

Figure 4. O-C times of all superhump maximum relative to the ephemeris in eqn (1).

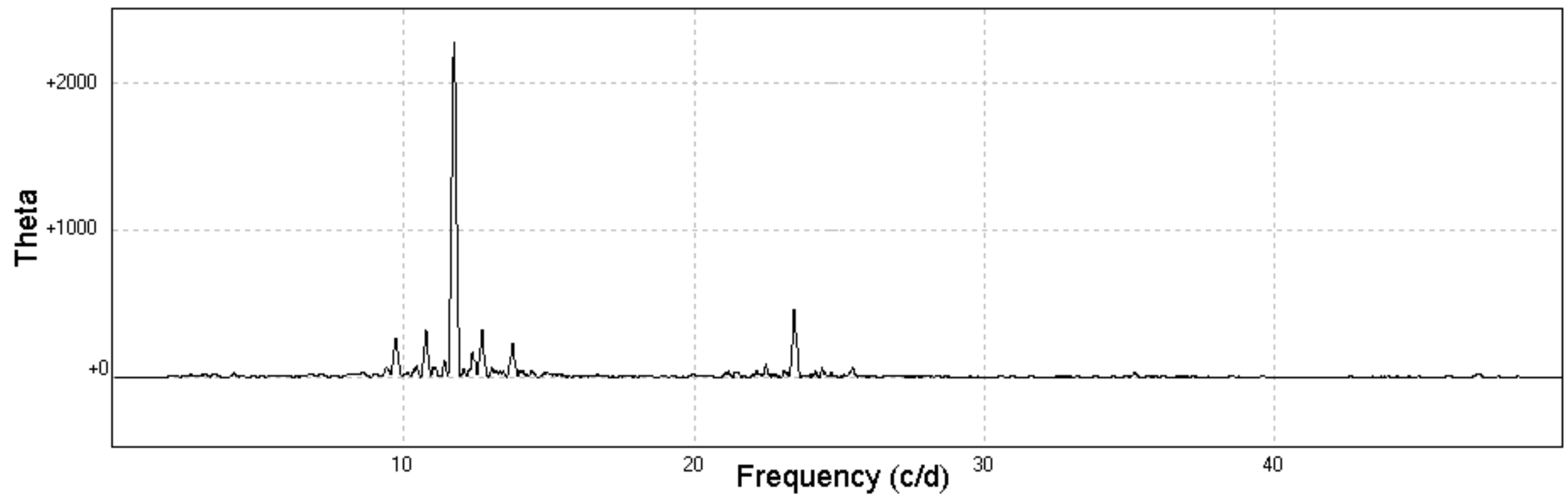

Figure 5. Power spectrum from Lomb-Scargle analysis.

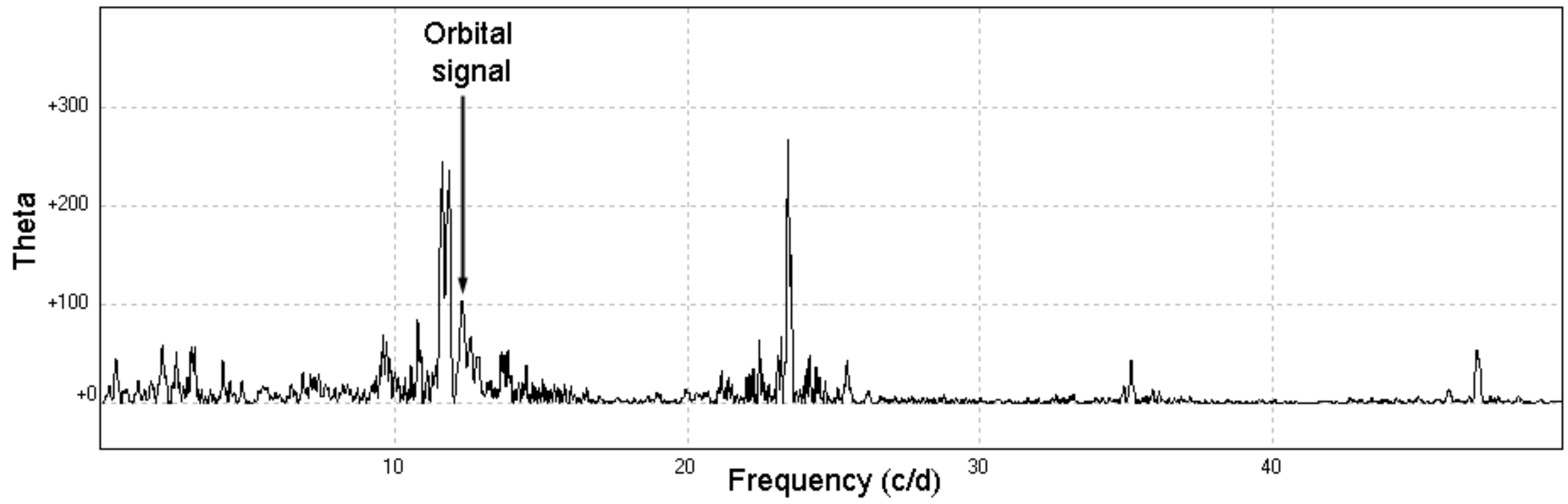

Figure 6. Power spectrum from Lomb-Scargle analysis after removing the superhump signal at 0.0853 d.

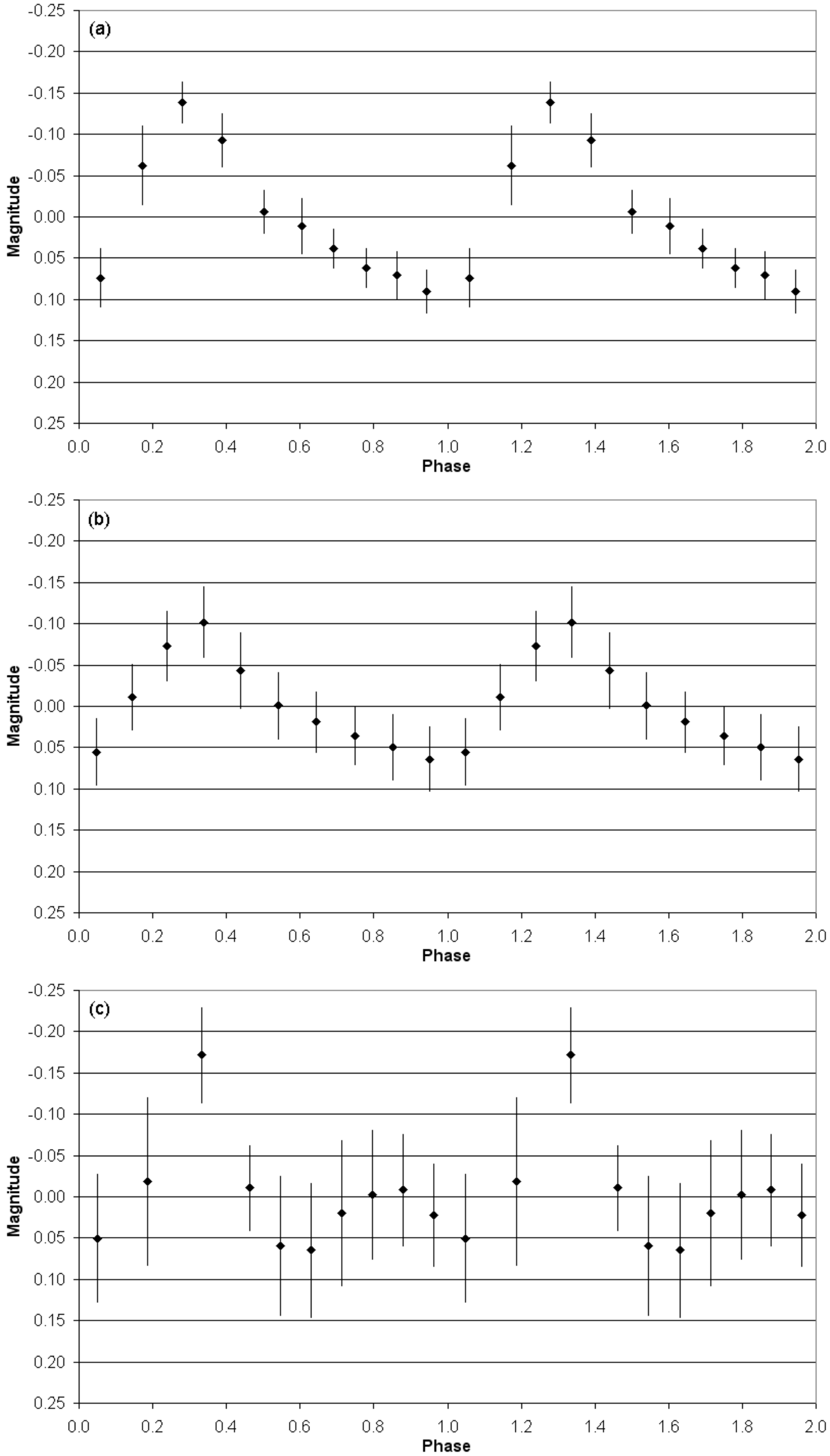

Figure 7. Superhump phase diagrams folded on the values of $P_{sh}$ in Table 6 for (a) cycles 0-15, (b) cycles 16-65 and (c) cycles 113-115.